# In-vivo interactions between tungsten microneedles and peripheral nerves

Pier Nicola Sergi[1], Winnie Jensen[2] , Silvestro Micera[1,3], Ken Yoshida[4]

[1] *BioRobotics Institute, Scuola Superiore Sant'Anna, Italy*

[2] *Department of Health Science and Technology, Aalborg University, Denmark*

[3] *Institute for Automation, Swiss Federal Institute of Technology, CH, Switzerland*

[4] *Biomedical Engineering Department, Indiana University-Purdue University Indianapolis, USA*

Corresponding Author

Dr. Pier Nicola Sergi

Scuola Superiore Sant'Anna

piazza Martiri della Libertà, 33 - 56127 Pisa - Italy

phone: +39050883135

fax: +39050883101

email: p.sergi@sssup.it

**Abstract**— Tungsten microneedles are currently used to insert neural electrodes into living peripheral nerves. However, the biomechanics underlying these procedures is not yet well characterized. For this reason, the aim of this work was to model the interactions between these microneedles and living peripheral nerves. A simple mathematical framework was especially provided to model both compression of the external layer of the nerve (epineurium) and the interactions resulting from penetration of the main shaft of the microneedle inside the living nerves. The instantaneous Young's modulus, compression force, the work needed to pierce the tissue, puncturing pressure, and the dynamic friction coefficient between the tungsten microneedles and living nerves were quantified starting from acute experiments, aiming to reproduce the physical environment of real implantations. Indeed, a better knowledge of the interactions between microneedles and peripheral nerves may be useful to improve the effectiveness of these insertion techniques, and could represent a key factor for designing robot-assisted procedures tailored for peripheral nerve insertion.

## 1.    Introduction

There are several medical procedures that involve the insertion of needles into biological tissues: needles are used to place radioactive seeds close to clusters of tumour cells (brachytherapy) [1],  to reach less accessible sites and extract biological specimens (biopsy) [2, 3], and to inoculate drugs with high accuracy inside the body (several types of injections) [4]. They are also  commonly used in microneurography [5, 6].

Since these kinds of procedures can be intrinsically difficult and can sometimes cause complications, also after intense training by surgeons, several studies were conducted to investigate the behaviour of needles during insertion in soft tissues. In these studies, the needles were modelled as linear elastic with geometric non linearities, [7-9] and simulations were validated using numerical models or phantom tissues. The work not only contributed to understanding these insertion procedures, but also provided a generalization to automate them. A particular class of insertion involves the use of microneedles which are used for placing  neural interfaces into peripheral nerves. In particular, ex-vivo insertion of these needles have been experimentally described in porcine [10] and rabbit [11] nerves.

Although the behaviour of peripheral nerves has already been studied in radial compression experiments aiming to simulate the application of cuff electrodes [12], the biomechanics of interaction between tungsten microneedles and living peripheral nerves has not been sufficiently characterized yet.

Indeed, this knowledge could be useful to improve the quality of standard insertion procedures, and could represent a key factor for  physically-based feedback aimed at monitoring robot-assisted procedures [4, 13] in peripheral nerve  puncture and insertion techniques.

Therefore, the aim of the work was to provide a simple theoretical framework for analysing the experimental results achieved during in-vivo insertion of tungsten microneedles. The framework was able to account for the main steps of the insertion process, considering on the one hand both the material of the microneedles and their tip geometry, and on the other the biomechanical-based response of the living nerves. In particular, the model was able to quantify not only piercing pressures, forces, and work, but also physical parameters, such as the instantaneous Young modulus of living peripheral nerves and the dynamic friction coefficient, characterizing tissue response and interaction with the tungsten microneedles.

This kind of data could be useful to improve interaction models accounting for complex and dissipative effects.

## 2.     Materials and methods

### 2.1     *Animal preparation*

In-situ acute experiments were performed on the peripheral nerves of three Danish Landrace pigs (~8-9 months old weighing about 45 Kg). All experimental procedures were approved by the Animal Experiments Inspectorate under the Danish Ministry of Justice.

The pigs were anesthetized using a mixture of Xylazine (Rompun®, 20 ml/l), Ketamine (Ketaminol®, 50 mg/ml), Butorphanol Tartrate (Turbugesic®, 10 mg/ml) and a combination of Tiletamin and Zolazepam (combined in Zoletil, 50 mg/ml). The animals were intubated and placed on a veterinary anaesthesia ventilator (model 2000, Hallowell EMC, USA) at 15 breaths/min. Anaesthesia was maintained using a 50-50% air/oxygen mixture with 1-1.2 % Isoflurane. The animals received 0.9% NaCl saline by continuous IV infusion through ear vein to prevent dehydration, and rocuronium bromide (Esmeron®, 10 mg/ml) and Fentanyl ("Hameln" 50 µg/ml) to provide analgesia throughout the entire experiment. The animals' heart rate and oxygen saturation were monitored throughout the experiment (Figure 1a).

### 2.2     *Experimental methods and data acquisition*

Surgical access was created in the upper forelimb to the ulnar and median nerves, and the site of insertion was chosen approximately 50 mm above the elbow. Precise implantations were performed at a constant velocity [14] of 2 mm/s, in the ulnar nerves. The compressive force was measured using a load cell (Sensotec Inc., Model 31/1435-02, max load 0.1 Kg, resolution 1.31 mV/g). A male fitting on the load cell was used to accept a female Luer lock onto which the needles were mounted (Figure 1b).

The nerves were elevated using a plastic platform with a 4 mm diameter hole (see Figure 1b) centered underneath them. The microneedle tips were perpendicularly placed slightly above the external surface of the nerves, and were then advanced into them by a motor-controlled (Maxon DC Motor 22-60-881, JVL Industri Elektronik A/S, DK) hydraulic micromanipulator (Narishige MMO-220) with a resulting movement resolution of 0.25 µm. After each insertion, the needle was repositioned in a different part of the nerve to avoid multiple insertion points in the same location.

Two groups of electro-sharpened tungsten microneedles, with different shaft diameters ($\varnothing$) and conical tips were used to perform 32 insertions: 19 and 13 trials were conducted using microneedles with $\varnothing$=75 µm and $\varnothing$=100 µm respectively. Moreover, all of them were assumed to have Young modulus and Poisson ratio of $E_n$=411 GPa, and $v_n = 0.28$ respectively.

Round tungsten rods (no insulation, 75 and 100 µm diameters, A-M Systems) were cut into 2 cm lengths and then manually electro-sharpened to create a sharp tip. The tips were sharpened by lowering one end into a 2N $KNO_3$ solution, placing a ~8 VAC potential across it, and using a large carbon counter electrode for 20-40 seconds to etch them. The tips were visually inspected through a microscope during and after the electrolyses procedure. Finally, the opening angle of the tip was measured following optical photomicrography (Figure 1c). All electrodes were cleaned in de-ionized water before implantation to remove any residue from etchant.

The microneedles were inserted according to a ramp-hold-ramp profile (Figure 1d), and the force and position signals (Figure 1e) were filtered before sampling (1st order lowpass filter, sampling frequency = 2.5 KHz, NI DAQ card-6204E). The sampled data were also filtered offline, to minimize superimposed physiological and instrumental noise. In particular, the relaxation phase was filtered using a $3^{rd}$ order Butterworth lowpass filter ($F_{3dB} = 0.325$ Hz), while the compression and insertion phases were filtered with a $3^{rd}$ order Butterworth lowpass filter ($F_{3dB} = 12.5$ Hz).

## 2.3    *Modelling of compression and puncture*

When the tip of a microneedle comes into contact with the outermost layer of the nerve (epineurium), the pressure increases (Figure 1e, phase A) and the layer deforms: the relation between these sets of quantities is governed by constitutive equations.

Since a nonlinear Kelvin model [15] was used to approximate the general behaviour of the living nerves, and the velocity of microneedles was constant for all phases (Figure 1e, phases A-B-D), the interaction forces were written as :

$$(1) \quad F(z, v\tau_s) = f_s(z) + k(z)v\tau_s \left[ 1 - \exp\left( -\frac{z}{v\tau_s} \right) \right]$$

where $z$ is the dimpling of the tissue (which, in this case, equals the microneedle tip displacement), $v=dz/dt$ is the velocity of the needle tip, $f_s(z)$ is a nonlinear force-displacement function and $k(z)$ a nonlinear spring connected with a nonlinear damper $b(z)$, with a characteristic relaxation time $\tau_s=b(z)/k(z)$. The exponential term of Equation (1) depends on the dimpling of the tissue before the puncture ($z$), as well as on the characteristic relaxation time of nerve ($\tau_s$) and piercing velocity ($v$). Since in this study $v\tau_s >> z$, the asymptotic expansion of Equation (1) with respect to $v\tau_s$ was used:

$$(1.1) \quad F(z, v\tau_s) = f_s(z) + k(z)v\tau_s \left[ 1 - \exp\left( -\frac{z}{v\tau_s} \right) \right] = f_s(z) + k(z)\left( z + O\left[ \frac{1}{v\tau_s} \right] \right)$$

and Equation (1) was approximated with Equation (1.2):

$$(1.2) \quad F(z) \cong f_s(z) + k(z)z$$

Equation (1.2) shows that the total force is independent of the viscous characteristics of the tissue for any choice of $f_s(z)$ and $k(z)$.

Above all, due to the small size of the needle tip (<< 100 μm) with respect to the nerve diameter (3-5 mm), this problem was treated as an indentation of an incompressible elastic tissue (i.e. the living nerve) performed with a conical indenter (i.e. the microneedle tip). Consequently, Equation (1.2), which implicitly relates forces and dimpling, was made explicit by merging the two terms and using Sneddon's approach [16, 17] together with the changes due to the real geometry of the conical tip [18]. Furthermore, the living peripheral nerves were considered a nonlinear elastic material, and the measured load $F(z)$ was written as:

$$(2) \quad F(z) = \frac{8\, tg(\alpha)}{3\pi} M(E,z)\, z[z + \overline{\psi}(\rho)]$$

where $E$ is the instantaneous Young's modulus of the nerve, $M(E,z) = E \exp(bz)$ models the nonlinear elasticity of the nerve [12] undergoing a finite deformation (indentation) and assumed incompressible [19], $\alpha$ is the half angle of the needle tip, $z$ is the dimpling of the nerve (or tip displacement), and $\overline{\psi}(\rho)$ is a function of the curvature radius of the tip, with the dimension of a length. According to [18], for both groups of microneedles, Equation (2.1) can be written as:

$$(2.1) \quad \overline{\psi}(\rho) = c_1 \rho^2 + c_2 \rho$$

where $c_1 = 1.5 \cdot 10^{-4}$ m$^{-1}$ and $c_2 = 1.17 \cdot 10^{-1}$ .

The half angles ($\alpha$) and the tip radii ($\rho$) of the microneedle tips were measured from digital pictures, and Equation (2) was used to fit experimental data (phase A) through non-linear optimization with parameter identification based on measured compression forces (quasi Newtonian algorithm, Scilab © INRIA ).

All experiments performed with microneedles with $\varnothing$=75 μm and $\varnothing$=100 μm were separately grouped to extract the mean values of the quantities ($E$, $b$) characterizing the tissue response.

To integrate this direct biomechanical approach, an indirect sensitivity analysis was performed to further investigate the degree of nonlinearity of the compression force. Let $q(z,n)$ be a polynomial function of $n$ order, as shown in Equation (2.2):

$$(2.2) \quad q(z,n) = \sum_{j=1}^{n} \gamma_n z^n$$

where $\gamma_n$ are the constant coefficients of each $n$-order term (the 0-order term is set identically to zero because the force of compression is zero for $z=0$). The set of experimental curves was fitted with these phenomenological functions by varying the maximum order of polynomial approximation ($n$-index) in order to evaluate both the residuals and $R^2$ index ($R^2$ is the square of the correlation between the response values and the predicted values). The order of $n$ for which the improvement of indices stops is the order of nonlinearity of Equation (2.2).

Recalling Equation (2), the global work applied to the nerve during compression up to the puncture was written as:

$$(3) \quad W = \int_{0}^{z_0} F(z)dz$$

where $z_0$ is the coordinate where the tissue was pierced. The nerve was punctured when a characteristic maximum pressure was exceeded, the surface of the nerve yielded and a crack opened in the tissue. At this stage, the central radius of the crack was assessed with the radius of the microneedle tip. During insertion, the size of the crack increased up to the external diameter of the main shaft of the microneedle: in the event of yielding, the theoretical pressure at the microneedle-nerve interface was roughly assessed by using the following equation [20]:

$$(4) \quad p(r) = \frac{1 - r_t/r}{\dfrac{1}{E_n}\left(1 - \nu_n\right) + \dfrac{3}{2E_{ts}(\varepsilon_r)}}$$

where $E_n$ and $\nu_n$ are the Young and Poisson modules of the needle, $r$ and $r_t$ are the radii of the needle shaft and of the needle tip, and $E_{ts}$ is the instantaneous Young modulus of the tissue.

When the initial micro crack expanded into a hole, the nonlinear behaviour of the tissue was modelled with the function $E_{ts}(r) = E \exp(\beta \varepsilon_r)$, where $\varepsilon_r = (r_n - r_t)/r_t$ and $\beta > 0$ represent an exponential constant modelling the radial strain hardening of the tissue [12]. Equation (4) is strictly valid for a pair of perfectly elastic materials and usually provides high pressure values when the $r/r_t$ ratio is high, as in this case.

Nevertheless, the maximum pressure that can be obtained at the interface is yielding pressure; viscous flows arise beyond this threshold which lower the local field of contact stresses. Equation (4) shows the need for yielding at the interface between the nerve and the microneedles but is not able to quantify it. To this aim, it could be useful to recall the behaviour of the nerve under the microneedle tip, before opening of the hole: the tissue yields and viscous flows arise before fracturing of the outermost layer (viscoelastic fracture).

As a consequence, the superficial yielding pressure was reasonably approximated with the piercing pressure:

(5) $p(r) = \overline{p}_y = \dfrac{F(z_0)}{A}$

where $A = \pi \left[ 2\rho^2 + (r+\rho)\sqrt{(r-\rho)^2 + L_1^{\,2}} \right]$ and $\rho$ is the radius of curvature of the microneedle tip, $r$ is the nominal diameter of the main shaft, and $L_1$ is the height of the conical region at the end of the needle [15,21]. To achieve a mean value of this parameter, digital pictures of microneedles were analysed and measured. It should be noted that Equation (5) provides the superficial yielding pressure, which differs from the mean pressure over the microneedle cross sectional area given by:

(5.1) $\overline{p}_c = 4F(z_0)/\pi d^2$

where $d$ is the nominal diameter of the microneedle.

A total force is applied on the microneedles after puncturing of the tissue, which derives from the sum of friction and a constant cutting force [14]. Above all, the total interaction force increases almost linearly during insertion together with the contact surface of the microneedle.

For the sake of simplicity, the friction was modelled as Coulombian, then the mean dynamic coefficient was written as:

$$(6) \quad \overline{\mu}_d = \frac{\overline{\tau}}{\overline{p}_y} \mathrm{sgn}(v)$$

where $\overline{\tau} = m/2\pi r$ is the mean shear stress on the lateral surface of the inserted microneedle, $m$ is the derivative respect to $z$ of the straight line fitting the total force, and sgn($v$) is the sign function of the microneedle velocity $v$. The piercing pressure and the slope of the insertion force are univocally related for each curve. As a consequence, each curve was separately analysed to quantify the friction coefficient avoiding errors coming from the use of mean values. Additionally, this approach allowed us to investigate the nature of the distribution of the dynamical friction coefficient among experiments.

## 3. Results

### 3.1 An elastic framework to describe interaction between nerve and microneedles

For insertion experiments performed with the first group of tungsten microneedles ($\varnothing$=75 μm) mean characteristic relaxation time and mean dimpling were respectively $\overline{\tau}_{s75} = 28.9558\ s$ and $\overline{z}_{075} = 2.6456\ mm$, while for experiments performed with the second group they were ($\varnothing$=100 μm), $\overline{\tau}_{s100} = 28.8158\ s$ and $\overline{z}_{0100} = 3.6369\ mm$. In both cases, $v\overline{\tau}_{s75} >> \overline{z}_{075}$ and $v\overline{\tau}_{s100} >> \overline{z}_{0100}$, then Equations (1.2) and (2) were used and the elastic framework was able to correctly describe the experimental results.

## 3.2 Geometry of microneedles tips and instantaneous response of the nerve

The half angles ($\alpha$) and the tip radii ($\rho$) of the microneedle tips for the first group were $\alpha = 10.25 \pm 0.5°$ and $\rho = 1.7 \pm 0.577$ μm, while for the second group they were $\alpha = 7.875 \pm 0.478°$ and $\rho = 2.48 \pm 0.349$ μm: these values were inserted in Equations (2) and (2.1) to fit data resulting from experimental indentations (phase A).

Figure 2 shows both experiments and fitting curves: Equation (2) fitted experimental data of the first group ($R^2 = 0.8591$) to the set { $\overline{E}_{75} = 22.2265$ kPa , $b_{75} = 0$, $\overline{\psi}(\rho)_{75} = 1.99 \cdot 10^{-4}$ $mm$ }, and also fitted experimental data of the second group ($R^2 = 0.9649$) to the set { $\overline{E}_{100} = 32.9663$ kPa , $b_{100} = 0$, $\overline{\psi}(\rho)_{100} = 3 \cdot 10^{-4}$ $mm$ }.

Furthermore, the indirect analysis of the compression clearly showed a quadratic nature of the phenomenon (see Figure 3a-f). The analysis of both residuals and $R^2$ indexes did not support this conclusion. Indeed, in both cases for $n \geq 2$ there was a superimposition of residuals and a constancy of $R^2$ as shown in Figure (3c-e).

However, the relevance analysis of the coefficients of fitting functions (Figure 3f) showed that only the linear and quadratic terms were different from zero for all polynomials with $n \geq 2$ and for both groups of microneedles.

## 3.3 Piercing and insertion features

Tissue dimpling and piercing was estimated with Equation (3). To this aim, the contact areas for both groups were calculated as $A_{75} = 0.03304 \pm 0.0038$ mm$^2$ and $A_{100} = 0.0663 \pm 0.0130$ mm$^2$, and Equation (3) led to $\overline{W}_{75} = 28.4912$ $\mu J$ and $\overline{W}_{100} = 60.8844$ $\mu J$. Moreover, Figure (4a) shows the box plots of both quantities $W_{75}$ and $W_{100}$. In particular, the lines inside the central boxes represent the medians $\widetilde{W}_{75} = 22.9210$ $\mu J$ and $\widetilde{W}_{100} = 54.0432$ $\mu J$, the upper and the lower lines of the boxes show the upper and lower quartiles, and the error bars include outliers. Figure (4b) shows the box plots of the piercing forces: the mean values were $\overline{F}_{p75} = 27.4301$ mN and $\overline{F}_{p100} = 54.1736$ mN, while the medians of the same quantities were $\tilde{F}_{p75} = 24.094$ mN and $\tilde{F}_{p100} = 51.4703$ mN.

Equation (4) provides a much larger theoretical interfacial pressure than the yielding pressure of any soft tissue. Indeed, for $\beta = 0.145$ and $\overline{E}_{75}$ or $\overline{E}_{100}$, the pressures were found to be

$\overline{p}_{75} \approx 110$ MPa and $\overline{p}_{100} \approx 162$ MPa, while the traction yield stress for a 1002A steel was 131

MPa. It should be noticed, according to [12], that a value of $\overline{\beta} = 4.03$ was found for a circular compression of rabbit sciatic nerves. Therefore, Equations (5) and (6) were used to approximate the yielding pressure and the dynamic friction coefficient.

Indeed, Figure (5) shows the box plots of the piercing pressures and dynamic friction coefficients for each specimen and for each kind of microneedle. The mean and median values of piercing force, yield contact pressure, pressure on the cross sectional area and dynamic friction coefficient for microneedles with $\varnothing = 75$ μm were: $\overline{p}_{y75} = 0.8302$ MPa, $\overline{p}_{c75} = 6.2121$ MPa,

$\overline{\mu}_{d75} = 0.1626$, and $\tilde{p}_{y75} = 0.7292$ MPa, $\tilde{p}_{c75} = 5.4565$ MPa, $\tilde{\mu}_{d75} = 0.1748$. Similarly, for the

$\varnothing = 100$ μm group, the same quantities were $\overline{p}_{y100} = 0.8171$ MPa, $\overline{p}_{c100} = 6.9010$ MPa,

$\overline{\mu}_{d100} = 0.1900$ and $\tilde{p}_{y100} = 0.7763$ MPa, $\tilde{p}_{c100} = 6.5567$ MPa, $\tilde{\mu}_{d100} = 0.1726$.

Finally, Figure (6) shows the quantile-quantile plots of distribution of the dynamic friction

coefficient among experimental trials for both groups of microneedles. In particular, the Shapiro-

Wilk test was performed to assess the degree of "normality" of distribution (R: A language and

environment for statistical computing. R Foundation for Statistical Computing, University of

Vienna). For the first group, this test resulted in W = 0.9311, p-value = 0.1623, while for the second group, W = 0.9452, p-value=0.5278, where W is the square of the correlation between experimental data and normal distribution, while the p-value is the probability of obtaining a test statistic at least as extreme as the one that was actually observed, assuming that the null hypothesis is true. In this case, the null hypothesis was that experimental data $(\mu_d)$ came from a normally distributed population.

## 4. Discussion

### 4.1 *A simple framework to model superficial interactions*

In this manuscript, the interactions between tungsten microneedles and the nerve were described using a simple mathematical framework. Specifically, the first phase of compression was modelled with an elastic indentation. The more the viscous effects were negligible in a standard Kelvin model, the more this simplification approached the exact solution. As previously shown, the viscous effects slightly influenced the response of the peripheral nerves according to the chosen velocity of puncture and insertion, which approximated that of surgical implants. In particular, viscous effects mainly affected the end part of indentation; furthermore, maximum difference with a fully elastic description was approximately 2% at the end of compression, while at the beginning it was negligible.

Moreover, the classical Sneddon's approach was not directly used, since this approach is strictly valid for an infinitely sharp conical indenter, while the real needles had tips with a finite radius of curvature. Therefore, a modification of the classical theory was used according to [12] to extract the instantaneous Young's modulus of the living peripheral nervous tissue.

In particular, the mean values of this parameter (22.2265 – 32.9663 kPa) were comparable with mean data from other compression experiments on rabbit nerves. Indeed, a Young modulus of 41.6±5 kPa [22] was found for in vitro unconfined compressions, and a value of 66.9±8 kPa [12] was found for in vivo circular compressions.

Furthermore, the achieved values were comparable with physical properties of other biological tissues and showed that the living peripheral nerves of pigs were as stiff as dermis (35 kPa). Again, their stiffness had the same order of magnitude of muscle stiffness (80 kPa) [23], and they were more compliant than articular cartilage [19].

In addition, although peripheral nerves were considered as nonlinear elastic (Equation (2)), the achieved values of $b$ ($1 \cdot 10^{-2} < b < 1 \cdot 10^{-3}$) showed a practically linear behaviour during localized compressions ($M(E_{75,100}, z) \approx \overline{E}_{75,100}$) with limited values of $z$. As a consequence, the Young modulus of living nerves was practically constant and equal to the instantaneous Young modulus, and for sake of simplicity the parameter $b$ was set to zero. This approach led to a polynomial (2°order) indentation law consistent with the indirect phenomenological analysis (Figure 3a-f) and with the approach given in [14].

Finally, this simple framework was considered correct since the characteristic relaxation time of the nerve multiplied by the velocity of indentation was greater than the characteristic deflection of the nerve surface. Since peripheral nerves may experience a limited dimpling under the microneedle tip, this theoretical framework is still valid and usable for all tissues having similar characteristic relaxation time and equal or higher insertion velocities. Nevertheless, viscoelastic effects have to be considered for low insertion velocities or for shorter characteristic relaxation time.

### 4.2 Piercing and insertion of microneedles into peripheral nerves: interaction features

Compression forces were used to gauge interactions between living nerves and microneedles. In particular, the achieved piercing forces were greater than those found in similar experiments with ex vivo porcine [10] and rabbit [11] nerves. This could be due to differences in experimental set up. In [10], indeed, the nerve was kept under axial tension in a saline bath, whereas in these experiments the nerves were kept in the air, and a perforated support was placed under them to ease piercing and the insertion of microneedles. However, in both cases, the parts of the nerve interacting with the microneedles were free to dimple under the tip forces.

Another reason for this was the greater stiffness of the living nerves with respect to similar ex-vivo specimens. Indeed, a dense microcirculation system inside the living nerves provides nourishment to all internal substructures, and pressurized blood flows in small veins and capillaries. Moreover, the nervous fascicles are filled with a physiological fluid under pressure. As a consequence, the global tissue response to localized indentation is affected by this internal pressurization. Furthermore, in ex-vivo specimens, the lack of internal pressure in fascicles, small veins and capillaries, and post-mortem autolysis, which softens the connective tissues, work together to lower the global response of the nerves to external mechanical stimuli. The difference between in vivo and ex vivo piercing forces was most likely due to the complex interaction of these two causes, which generally seems to be valid for other tissues also [25-27].

Again, it could be worth pointing out the differences between the mean yielding pressures and the mean cross sectional pressures. Indeed, for the analysed groups of microneedles the mean cross sectional pressures were ~7.4825 and ~8.4458 times greater than the superficial yielding pressure. In particular, since microneedles have to bear the maximum piercing force without any buckling effect [10-11,28], the cross sectional pressure is relevant for designing both geometry and materials.

The previous difference was due to the procedure used for estimating the contact area. In this case, the area was the conical area of the tip in contact with the dimpled tissue, according to [21]. This choice led to a more close assessment of the force for the surface unit of the dimpled tissue. Since the tip of the microneedle sank into the surrounding (soft) tissue before compression, the contact area was assumed to be constant for the whole compression.

Equation (4) was used to assess the interfacial pressure between the main shaft of the microneedles and the living peripheral nerves. Since this approach exactly assesses the interference pressure between two elastic materials, the Young's moduli of tungsten and nerves were used in this work. Above all, as previously discussed, a microcirculation system nourishes the inner part of the nerve and the fascicles are filled with endoneural fluid: since the tissue is soaked in biological fluids, the incompressibility ($\nu \approx 0.5$) [19] of the whole nerve is acceptable and Equations (2),(4) were used in these simplified forms. Furthermore, the stiffness of the microneedles was assumed to be magnitudes higher than that of the living peripheral nerves: experiments confirmed this assumption ($E_n \gg \overline{E}_{75,100}$) allowing Equations (2) and (4) to be independent from the microneedle material. Furthermore, Equation (4) only showed the need for tissue yielding around the main shaft of the microneedle. Equation (5) was used, therefore, to assess this pressure because compression and yielding occur simultaneously just before piercing. Therefore, this procedure allowed us to assess a quantity which is difficult to explicitly achieve in experiments.

Finally, the interaction between peripheral nerves and microneedles was roughly modelled using a Coulombian dynamic friction model. This was a simplification of more complex phenomena that arose between the needle shaft and the surrounding tissue. Nevertheless, this simple model was able to catch the main macroscopic features of this phenomenon, giving physically reasonable values for the dynamic friction coefficient between biological and conventional materials, according to [29]. In addition, a statistical analysis of data was provided to investigate the course of the distribution of the dynamic friction coefficients. First, a rough analysis of the absolute differences between mean and median values was performed resulting in $1.22 \cdot 10^{-2}$ (~7.5% ) for the first group and $1.74 \cdot 10^{-2}$ (~10%) for the second. Although these values revealed a similarity, this kind of analysis was not able to investigate the global course of the distribution. Therefore, quantile-quantile plots were created and the Shapiro-Wilks normality test was performed. The plots showed an almost normal distribution, while the test provided high correlation indexes (W>0.9) and p-values>0.05 for both groups. Indeed, experimental data were close to the straight line (normal distribution) and the null hypothesis was accepted in the standard case of 5% of significance level.

## 5.     Conclusions and future works

Insertion experiments together with a simple mathematical framework were provided to investigate in vivo interaction between tungsten microneedles and ulnar porcine nerves. This approach allowed us to quantitatively assess the main physical quantities from which the experimental curves depend: the instantaneous Young's modulus, compression force, the work needed to pierce the tissue, puncturing pressure, and the dynamic friction coefficient between tungsten microneedles and living nerves were quantified. Specifically, the distribution of the dynamic friction coefficient was found to be approximately normal. This study could provide useful quantities for modelling and automating [13,30] peripheral nerve insertion tasks. Above all, this framework can be extended to all cases in which tissue dimpling is smaller than the product between characteristic relaxation time and piercing velocity: for long relaxation time coupled with quite low insertion velocity, or short relaxation time coupled with high insertion velocity.

Future work will extend the experimental trials to other living nerves featuring different kinds of needles and diverse combinations of relaxation time and piercing velocity, to explicitly account for viscoelastic effects.

## Conflict of interest

None

## Acknowledgements

The authors thank Dr. Jacopo Carpaneto for his valuable help. This work was supported in part by the European Union (EU) within the TIME Project (FP7-ICT –2007-224012, Transverse, Intrafascicular Multichannel Electrode system for induction of sensation and treatment of phantom limb pain in amputees).

## Figure captions

Fig. 1. (a) General view of experimental setup. (b) Magnification of tungsten microneedle and peripheral nerve with support. (c) Microneedle tip geometry, the measures are in degrees and micrometres. (d) The ramp-old-ramp profile of insertion-rest-extraction of the microneedle. (e) A characteristic curve (force vs. time), in which different phases are underlined. **A:** Compression phase (from beginning to point α). **α:** Point of puncture, where characteristic discontinuity in the compression force can be seen. **B:** insertion phase. **β:** End of the insertion phase, where the movement of the microneedle comes to a stop inside the tissue. **C:** Relaxation phase. **γ:** Retraction point, where the needle begins to be retracted from the tissue; **D** retraction phase. **δ:** Extraction of the microneedle from the tissue. The negative values of the forces (traction) are due to binding of the tissue on the external surface of the microneedle.

Fig. 2. Force-displacement curves during the phase of piercing. This phase can be theoretically described as an indentation. (a) Experimental (light grey) indentations with a ∅=75 µm tungsten microneedle, theoretical fitting curve (solid black) (b) Experimental (light grey) indentations with a ∅=100 µm tungsten microneedle, theoretical fitting curve (solid black).

Fig. 3. (a,b) Polynomial functions fitting experimental curves of compression with a $\phi$=75 µm (a) and a $\phi$=100 µm (b) microneedle. All functions with $n \geq 2$ perform the same fit and are shown superimposed to the solid black curve. (c,d) Residuals of polynomial fits of experimental data achieved with a $\phi$=75 µm (c) and $\phi$=100 µm. (d) for $n \geq 2$ residuals are exactly superimposed to the black area. (e) Changes of $R^2$ index varying the $n$-index (max degree of polynomial function and order of nonlinearity of the compression) for both sets of experiments. For $n \geq 2$ $R^2$ index remains constant, and does not allow direct assessment of the $n$-order of compression. (f) Relevance analysis of constant coefficients of all fitting functions: for both sets of experiments only the linear ($\gamma_1$) and quadratic ($\gamma_2$) terms are different from zero (light grey=0, dark grey>grey>0, white=not allowed), revealing a quadratic core of the compression phenomenon. For sake of simplicity only four powers are shown.

Fig. 4. (a) Box plots of work to reach the nerve puncture for both groups of experiments: the lines inside the central boxes represent the medians $\widetilde{W}_{75} = 22.9210 \ \mu J$ and $\widetilde{W}_{100} = 54.0432 \ \mu J$, the upper and the lower lines of boxes show the upper and the lower quartiles, the error bars include outliers. (b) Box plots of force for piercing the nerves using different diameters of microneedles: central lines represent the medians $\widetilde{F}_{p75} = 24.094 \ \text{mN}$ and $\widetilde{F}_{p100} = 51.4703 \ \text{mN}$.

Fig. 5. (a) Box plots of piercing pressure at puncture for different diameter of microneedles: central lines represent the medians $\widetilde{p}_{y75}$=0.7292 MPa and $\widetilde{p}_{y100}$=0.7763 MPa (b) Dynamic coefficient of friction during insertion into living nerves for both diameter of microneedles: central lines represent the medians $\widetilde{\mu}_{d75} = 0.1748$ and $\widetilde{\mu}_{d100} = 0.1726$ .

Fig. 6. Quantile-quantile plots of dynamic friction coefficients achieved with microneedles of $\varnothing$=75 mm (a) and $\varnothing$=100 mm (b). In both cases the normal distribution (straight bold line) is highly consistent with experimental values (dotted lines show 95 % confidence bounds).

FIGURE 1

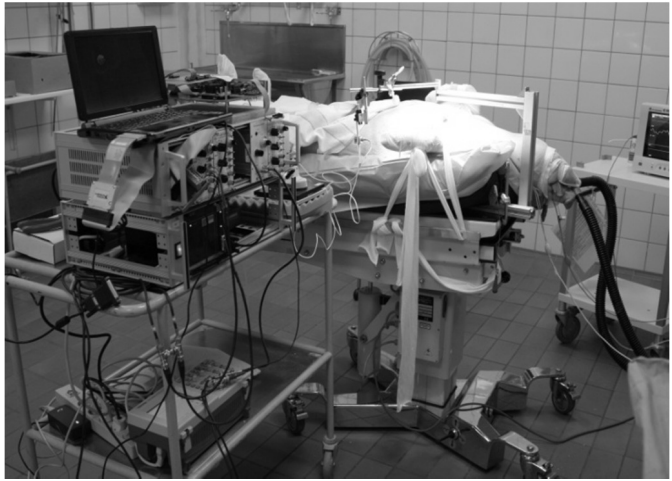

(a)

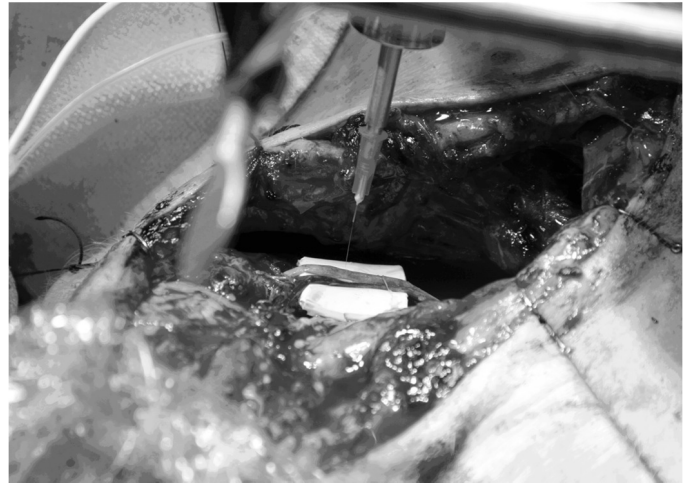

(b)

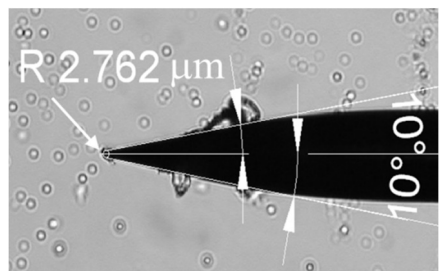

(c)

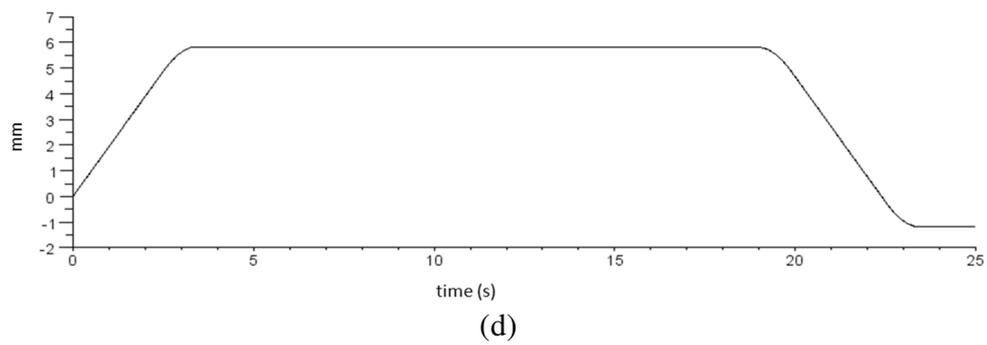

(d)

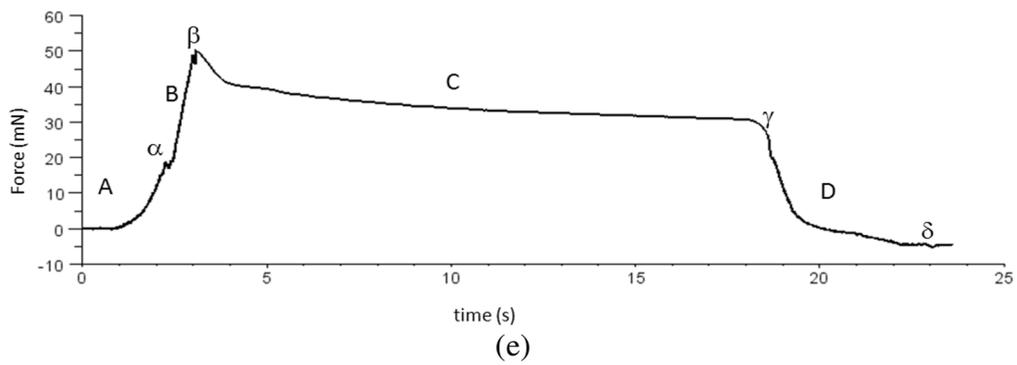

(e)

FIGURE 2

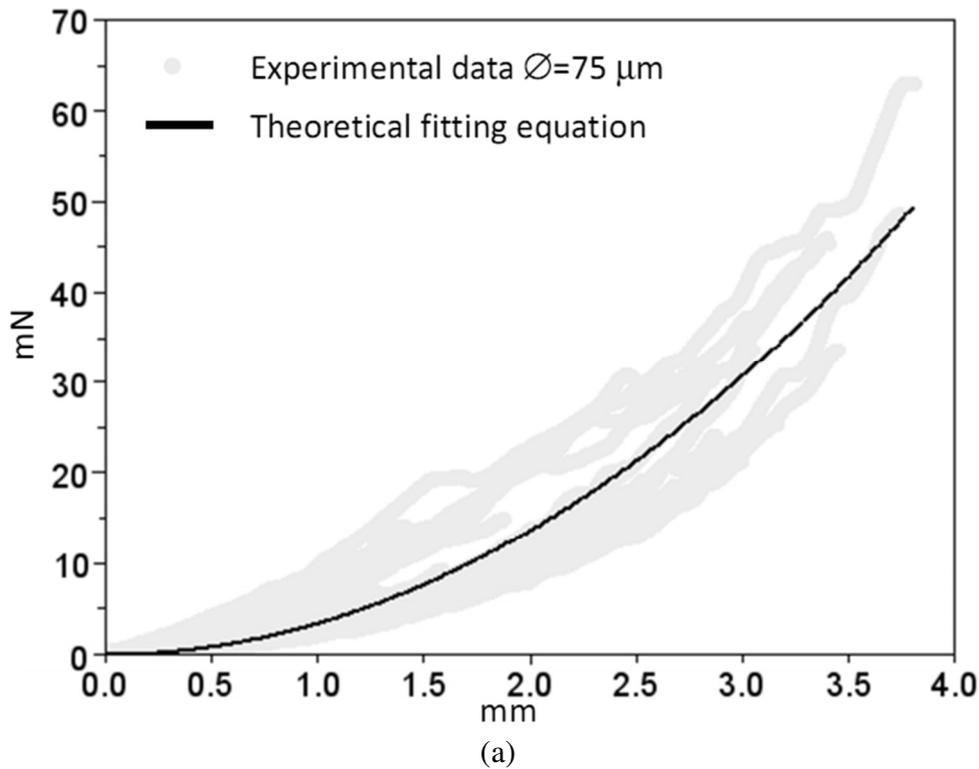

(a)

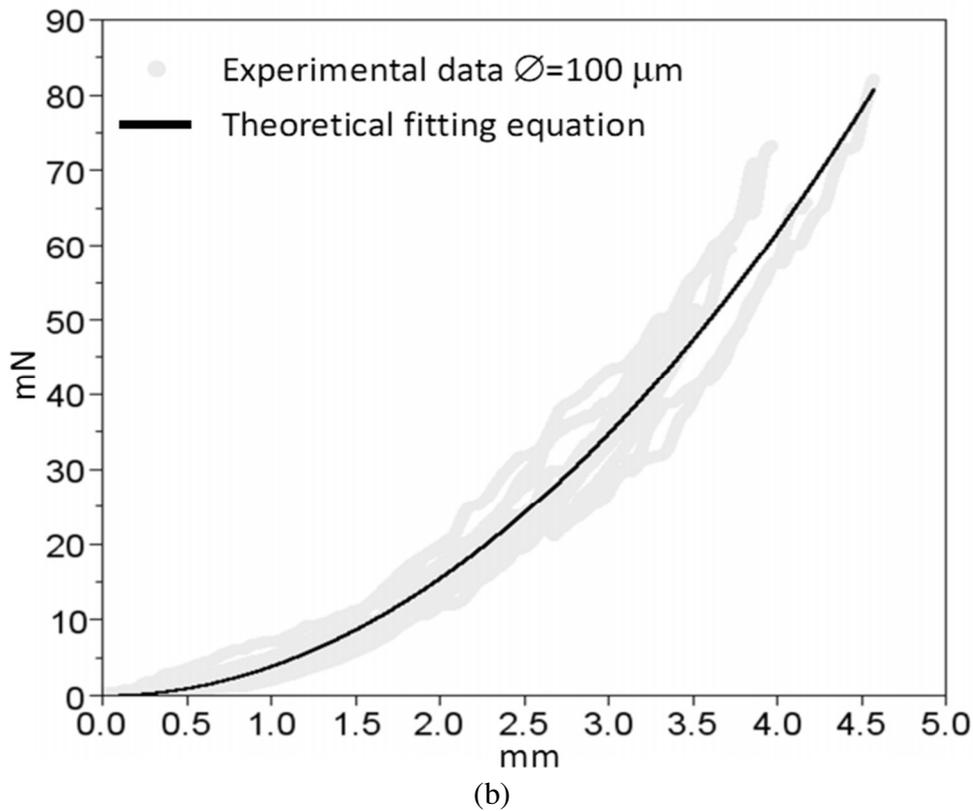

(b)

FIGURE 3

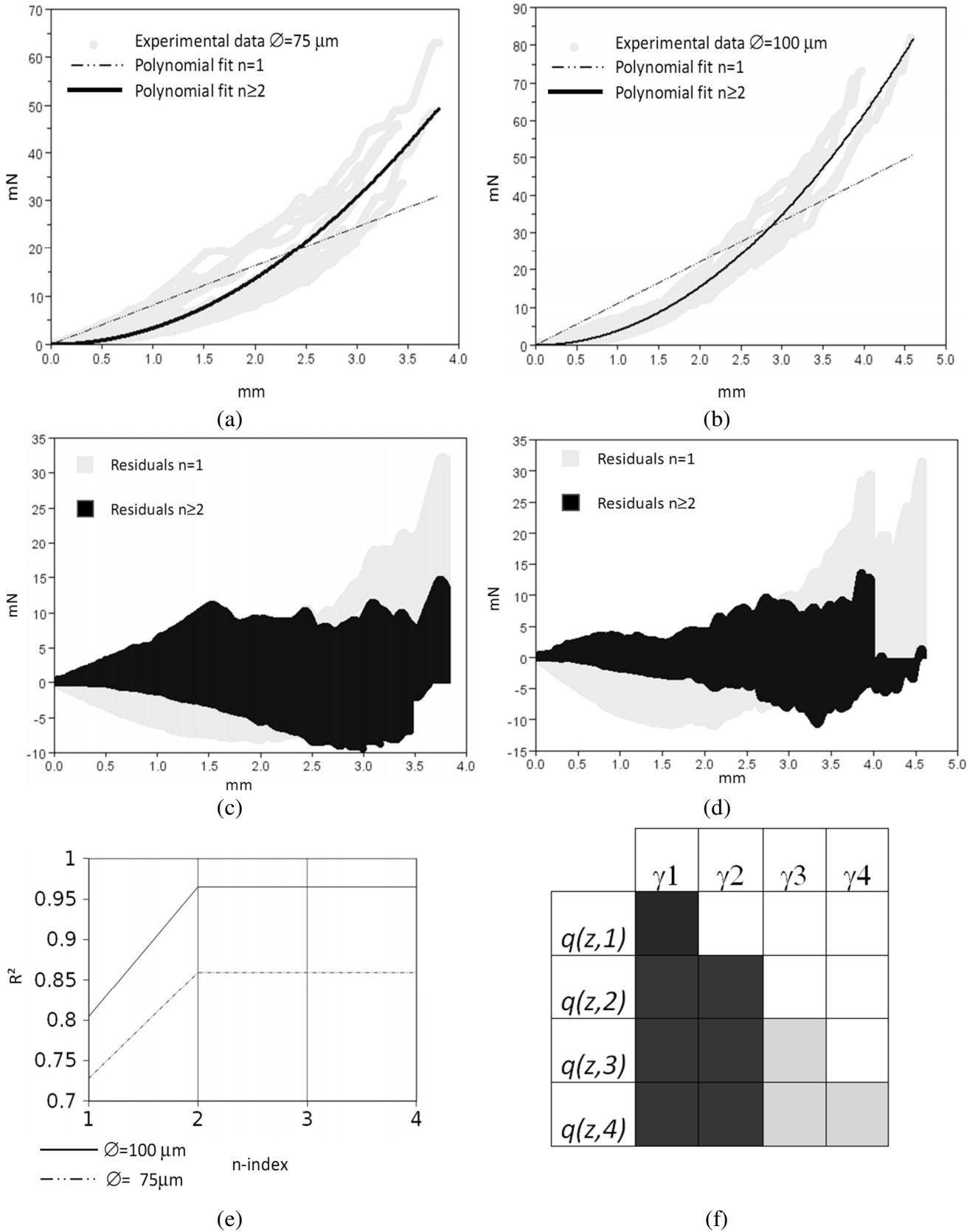

(a)  (b)

(c)  (d)

(e)  (f)

FIGURE 4

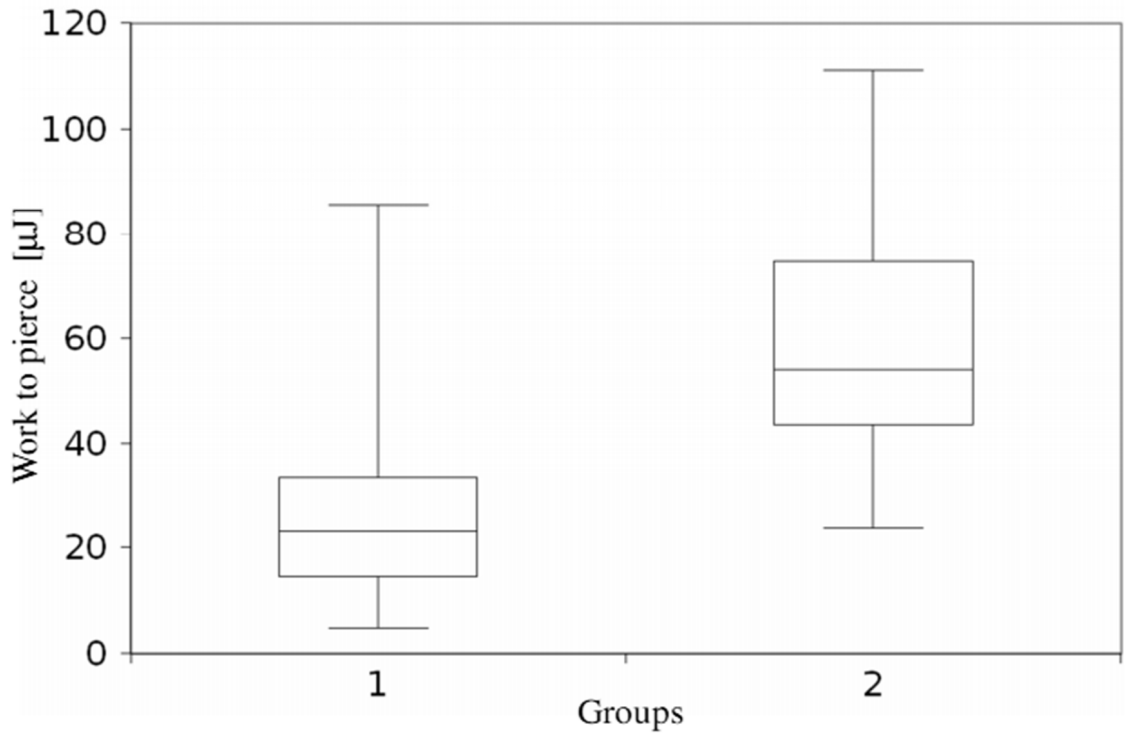

(a)

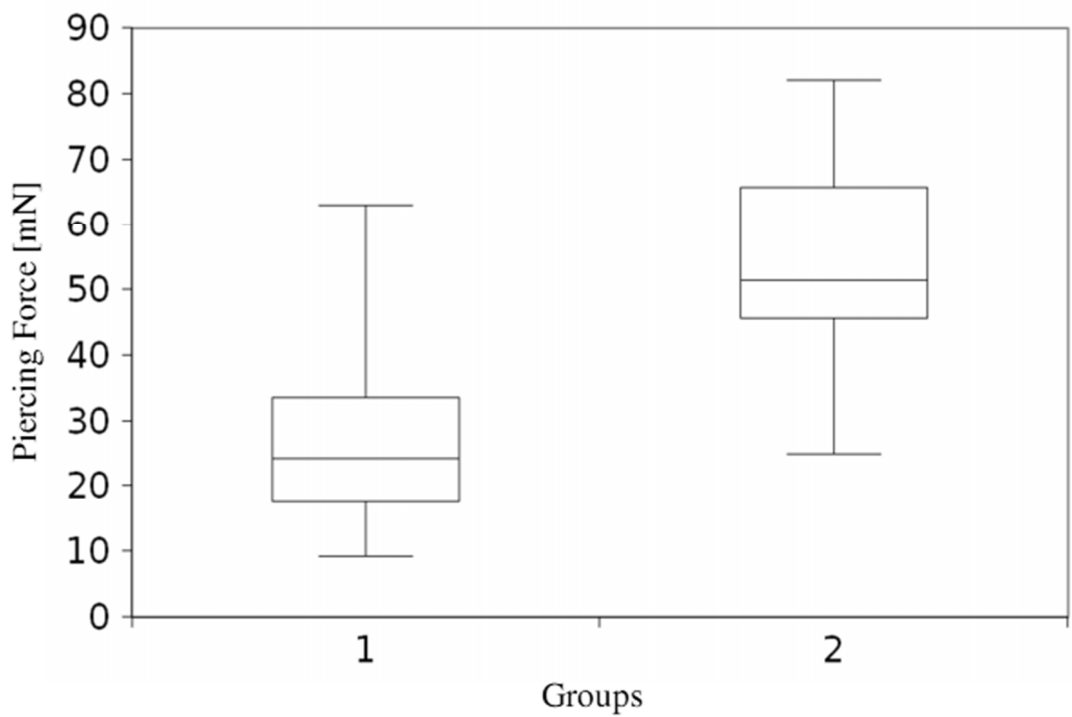

(b)

FIGURE 5

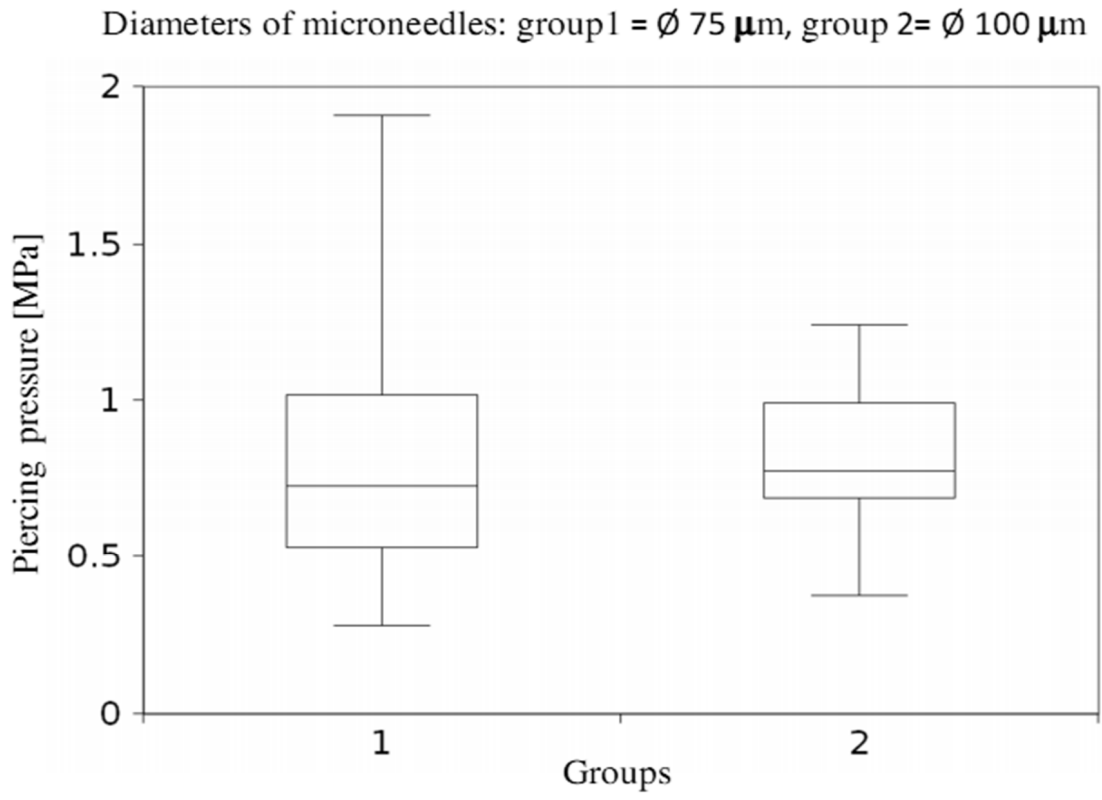

(a)

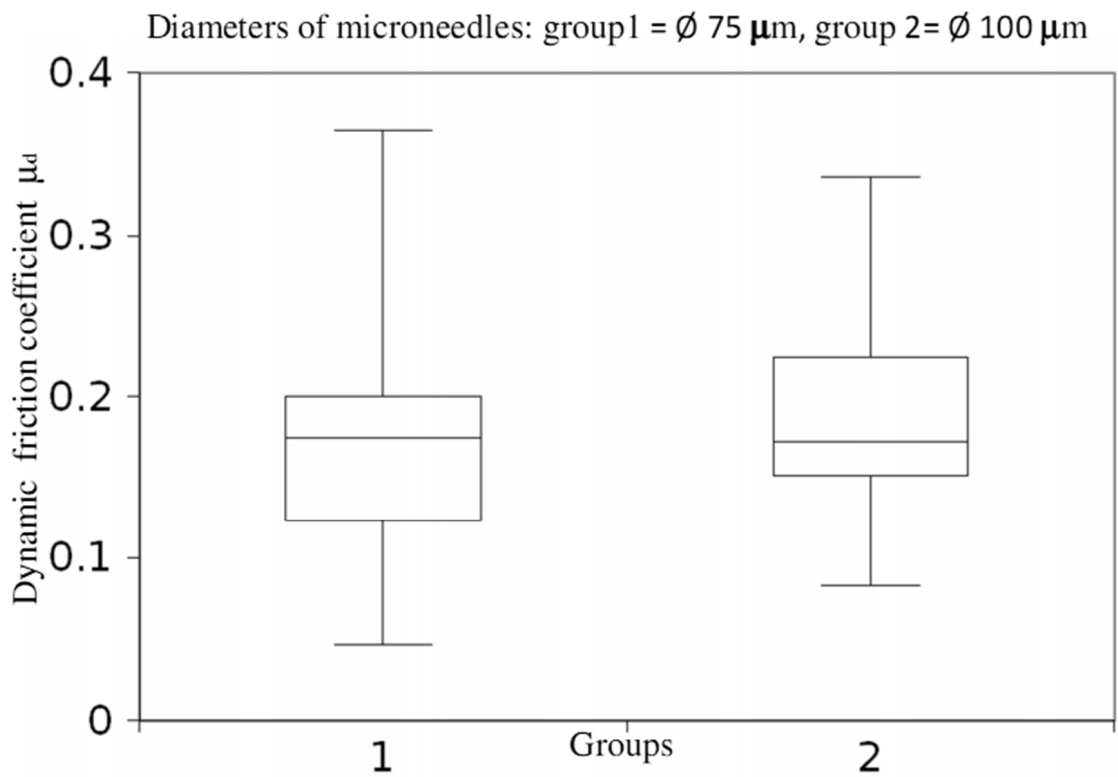

(b)

FIGURE 6

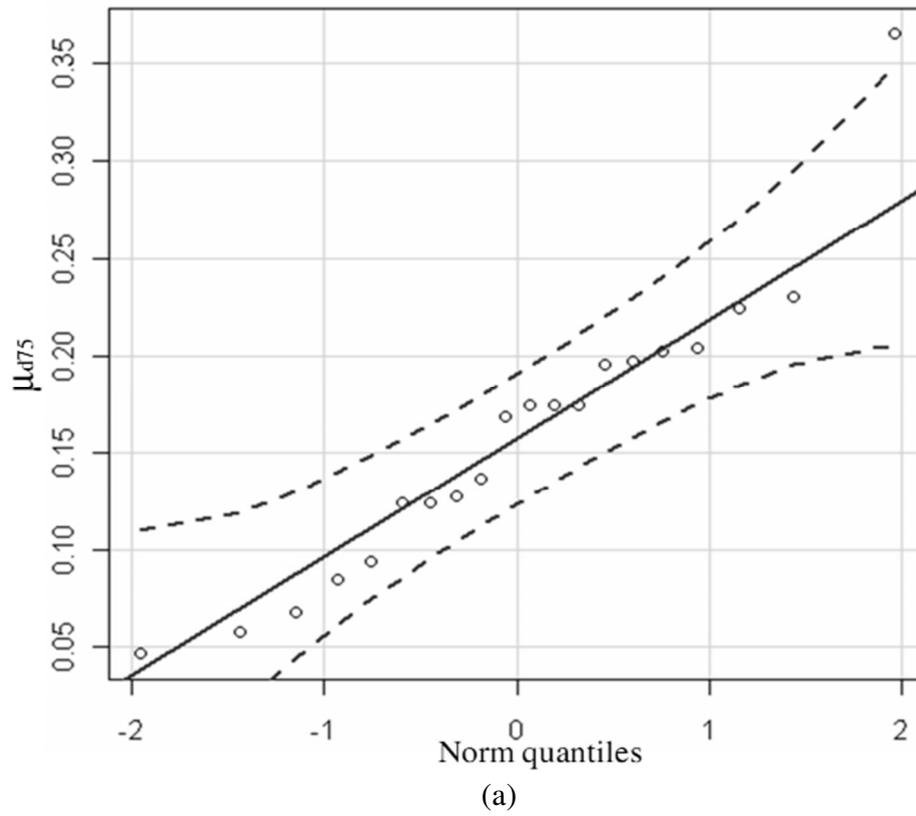

(a)

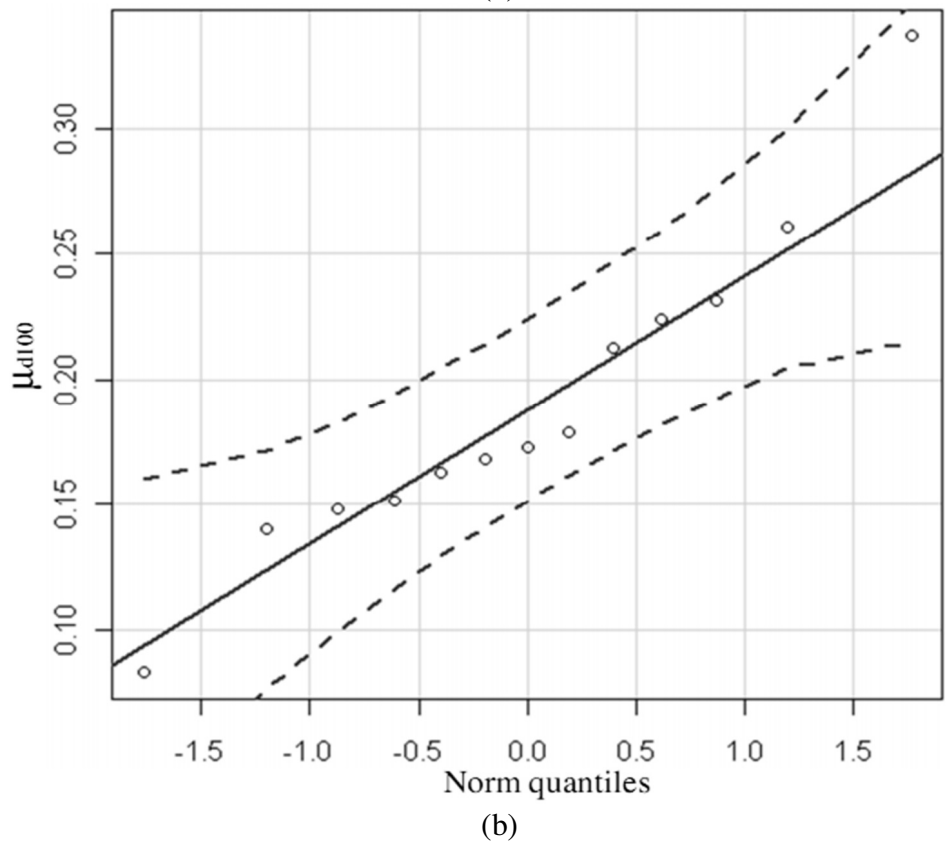

(b)